# High Energy Gamma Rays from Protons Hitting Compact Objects

J. Barbieri
NAWC-WD, China Lake, CA 93555
G. Chapline
Lawrence Livermore National Laboratory, Livermore, CA 94550

**Abstract**
In a previous paper the spectrum of positrons produced by matter initially at rest falling onto a massive compact object was calculated. In this paper this calculation is generalized to obtain both the spectrum of in-flight positron annihilation and $\pi^0$ decay gamma rays produced when protons with a cosmic ray-like spectrum hit the surface. The resulting $\pi^0$ decay $\gamma$ ray spectrum reflects the high energy proton energy spectrum, and is largely independent of the mass of the compact object. One notable prediction for all compact objects is a dip in the spectrum below 70 MeV. As applied to a $10^6$ solar mass massive compact object near to the center of our galaxy, our theory shows promise for explaining the $\gamma$ rays coming from the galactic center as observed by both the Compton satellite and HESS ground based array. In particular, if one uses the total intensity of 511 KeV positron annihilation radiation from the vicinity of the galactic center to fix the proton flux incident on the surface of the SgA* compact object and the spectral index recommended by Liu et. al. for stochastically accelerated protons, then while our theory underestimates the intensity around 1 GeV, it does explain both the overall spectral shape and intensity of $\gamma$ rays coming from the galactic center over the whole range from MeV to TeV energies.

## 1. Introduction

Over the past few years a new picture for compact astrophysical objects has emerged in which the interior space-time of the compact object resembles flat space-time with a large vacuum energy [5,13]. In this picture there are no space-time singularities in the interior of the compact object and there is no event horizon surface preventing information from escaping from the interior. Instead, compact objects have a real physical surface where, in sharp contrast with the celebrated prediction of classical general relativity that particles fall freely through an event horizon, one finds that depending on their energy particles can either fall into the interior and then reappear or decay into other particles as they pass through the quantum critical layer. The properties of this quantum critical layer should be quite analogous to the properties of condensed matter systems near to a quantum critical point. These critical properties are independent of the particular system, and this allows us to make definitive predictions for what happens to elementary particles when they encounter the surface of a compact astrophysical object. In particular, it is predicted that nucleons will decay into leptons and gamma rays when they fall onto the surface of a compact object [6]. In a previous paper [3] we calculated the spectrum of positrons that would be produced by matter initially at rest falling onto the surface of the compact object. These positrons can produce γ rays via annihilation with ambient electrons, and indeed this might be the explanation for the 511 KeV radiation coming from the center of our galaxy [10,11]. In this paper we extend our calculation of the positron spectrum resulting from nucleons at rest falling onto a compact object to the case of high energy protons with a cosmic ray-like spectrum hitting the object. In particular we report a calculation of spectrum of $\pi^0$ decay γ rays resulting from nucleon decay.

In the condensate vacuum ('dark energy star") model for compact objects introduced in [5] massless particles encountering the surface of a compact object see a quantum critical shell with thickness $z^* = R\sqrt{(\hbar\omega/2mc^2)}$ where $\hbar\omega$ is the energy of the massless particle, R is the Schwarzschild radius, and $m$ is a mass on the order of the Planck mass. When ordinary elementary particles such as photons or quarks enter this quantum critical layer they morph into particles with a large mass and have 4-point interactions characteristic of quantum criticality. One consequence is that all elementary particles can undergo 3-particle decays if their momentum before falling onto the compact object exceeds a threshold on the order $Q_0 = 100$ MeV/c $\sqrt{(M_o/M)}$, where $M$ is the mass of the compact object and $M_o$ is a solar mass. Elementary particles with momenta much less than this cut-off will pass freely through the critical surface, follow diverging geodesics in the interior of the condensate star, and eventually re-emerge through the surface. However, elementary particles whose momentum $\hbar k$ exceeds $Q_0$ will tend to decay as they pass through the quantum critical surface layer. It can be shown that the spectrum of decay products will have the universal form [5]:

$$\frac{dN}{d\Omega d\omega'} = \frac{27}{16\pi^2 \omega'}\left[3\frac{\omega'}{\omega}\left(1 - 3\frac{\omega'}{\omega} - 2\sqrt{\frac{\omega'}{\omega}}\cos\theta\right)\right]^{1/2}, \qquad (1)$$

where $\hbar\omega$ is the initial energy of the elementary particle that is incident on the compact object. $\hbar\omega'$ is the energy of decay product as seen by a distant observer, and θ is the angle of emission of the decay product with respect to the normal to the surface (backward emission). Because of the geometry of geodesics near to an event horizon the decay products that actually escape to infinity from the critical surface will be concentrated in a direction perpendicular to the surface, which means the emissivity of the surface is less than one.

It happens that the quarks and gluons inside nucleons typically have energies that exceed 100 MeV, so their their momenta will exceed $Q_0$ for essentially all compact astrophysical objects of interest. Therefore a generic prediction of the condensate model for compact astrophysical objects is that ordinary nucleons will disappear when they encounter the surface of the compact object [6]. In grand unification schemes for elementary particles such as the Georgi-Glashow SU(5) model [8] nucleon disappearance will proceed via the baryon number violating reactions:

$$u(2/3) \rightarrow e(+1) + \bar{u}(-2/3) + \bar{d}(1/3) \qquad (2a)$$

$$d(-1/3) \rightarrow e(+1) + 2\bar{u}(-2/3) \qquad (2b)$$

where $u$ and $d$ are the "up" and "down" quarks found inside protons and neutrons. In the Georgi-Glashow model nucleon decay will primarily yield positrons and mesons. Under ordinary circumstances nucleon decay is highly suppressed by the very large masses of the intermediate bosons associated with baryon number violation and "twist factors", and in fact has never been observed in the laboratory. However, for nucleons falling onto the surface of a compact object the energies of the constituent quarks and gluons will approach the "grand unification" scale where all allowed elementary particles interactions have approximately the same strength. Therefore baryon number violating processes are expected to be just as important on the surface of a compact object as the ordinary elementary particle processes studied in conventional accelerator experiments. In order to actually calculate the spectrum of decay products produced on the surface of a compact object, one must use the Altarelli-Parisi equations to find the distribution of momenta for the constituent quarks and gluons inside the nucleon. The resulting positron spectrum for nucleons initially at rest falling onto a massive compact object was calculated in a previous paper [3]. As it happens an excess of 511 keV annihilation radiation from positrons has been observed coming from the general vicinity of the center of our galaxy[10,, 11, 16], and to date there is no generally accepted conventional explanation for the origin these positrons. The absence of in-flight annihilation γ rays from the galactic center shows that the positrons producing the 511 keV radiation could not have been more energetic than a few MeV [4]. This is completely consistent with our calculated nucleon decay spectrum (see [3]). Thus the observations of positron annihilation γ rays from the vicinity of the center of our galaxy would appear to provide dramatic support for the idea that nucleons decay when they fall onto the surface of a compact object.

High energy protons, for example those in interstellar cosmic rays, will also decay when they hit the surface of a compact object, and produce positrons and hadronic showers. However, if the intensity of the primary cosmic rays as observed near to the earth is any guide, the product of the cross section of a compact object, $27\pi(GM/c^2)^2$, and the flux of interstellar cosmic rays would be far too small to explain either the positrons or γ rays coming from the galactic center. On the other hand, it has recently been pointed out that in general one expects there will be turbulent magnetic fields near to compact objects and these turbulent magnetic fields can accelerate protons to high energies (Liu et. al. 2006). In fact, as we will argue below this mechanism apparently can provide a flux of high energy protons that is sufficient to explain the γ rays coming from the direction of Sg A*, as observed by both the Compton satellite and ground based HESS telescope array.

## 2. Calculation

The hadronic jets resulting from the encounter of energetic protons with the quantum critical surface of a compact object will contain mesons, which immediately leads to γ rays coming from the surface as a result of the decay of $\pi^0$ mesons. The kinematics of this γ ray production from nucleon decay *are* somewhat different from positron production in that the positrons that we see must have been emitted in the backward direction (θ=0), whereas consideration of the angle-dependence of the fluorescence spectrum (1) and solid angle factors suggest that the $\pi^0$ decay γ rays that we see come mainly from hadronic jets emitted at angles close to tangential to the surface of the compact object. Actually emission of high energy gamma rays from jets directed inward (θ>π/2) is suppressed because the anti-quark is rapidly thermalized. The spectrum of γ rays resulting from $\pi^0$ decay can be estimated by convoluting the fluorescence spectrum (1) with the Altarelli-Parisi distribution for the momenta of the quarks and gluons inside the nucleons hitting the surface. The fluorescence spectrum (1) for θ=π/2 is peaked around $\omega'/\omega = 1/6$, with a maximum value of $\omega'/\omega$ of 1/3.

For the purposes of calculating the spectrum of $\pi^0$ decay γ rays we will adopt a "mini-jet" model where we assume that the hadronic jet produced by the decay anti-quark consists of a single $\pi^0$ whose momentum is equal to the momentum $\hbar\omega'/c$ of the decay anti-quark. That is we will set $\hbar\omega' = m_\pi c^2 \beta/\sqrt{(1-\beta^2)}$, where cβ is the $\pi^0$ velocity. The cross-section (1) and solid angle factors favor the production of decay anti-quarks near to θ=π/2, and hence the mini-jet $\pi^0$ s are produced predominately parallel to the surface of the compact object. Taking into account the transverse Doppler shift of the $\pi^0$ decay γ ray this allows us to identify the γ ray energy observed by a distant observer as

$$E_\gamma = \frac{\hbar\omega'}{2\beta}. \tag{3}$$

For small momenta this energy approaches 70 MeV. The spectrum for $\pi^0$ decay γ rays escaping to infinity will be given by

$$\frac{dN_\gamma}{dE_\gamma} = \eta \frac{1}{\beta} \int_{\omega_{min}}^{\infty} \frac{d\omega}{\omega} F(\omega',\omega) \int_{\omega}^{\infty} \frac{dN_p}{dE_p} \frac{q(x,Q^2)}{E_p} dE_p, \tag{4}$$

where $\omega_{min}$ is either $3\omega'$ or the cutoff frequency $\omega_c \equiv Q_0 c/\hbar$ depending on which is larger, $x=\hbar\omega/E_p$, $dN_p/dE_p$ is the proton spectrum, and η is the fraction of the $\pi^0$ decay γ rays that escape from the surface to infinity. $q(x,Q^2)$ is a normalized Altarelli-Parisi distribution for the quark momenta inside a nucleon and $F(\omega',\omega)$ is the normalized fluorescence spectrum. According to Liu et. al. [12] the spectrum of protons stochastically accelerated by magnetic fields near to a compact object will have the form

$$\frac{dN_p}{dE}(E) = F_o(E_0/E)^\alpha. \tag{5}$$

By changing the integration variables in eq. (4) to x and $\omega'/\omega$ one can show that for a proton spectrum of the form (5) the spectrum (4) can be written in the form:

$$\frac{dN_\gamma}{dE_\gamma} = \eta \frac{F_0}{\beta} < (\frac{\omega'}{\omega})^{\alpha-1} > < (\frac{\omega}{E_p})^{\alpha-1} > (\frac{E_0}{E\gamma})^\alpha \qquad (6)$$

where the average values of $\omega'/\omega$ and $\omega/E_p$ are calculated using $F(\omega',\omega)$ and the Altarelli-Parisi function $q(x)$ respectively. As is immediately evident from eq. (6) if the proton spectrum is a power law, then except near to 70 MeV the gamma ray spectrum will also be a power law with the same slope. In Fig 1 we show our calculated $\pi^0$ decay γ ray spectrum using the spectral index α =2.2 recommended by Liu et al. [12]. Amusingly this is same spectral index observed for primary cosmic rays near to the earth [7]. The Altarelli-Parisi $q(x)$ is the same one we previously used for positron emission, except that now the initial proton energy is interpreted as the energy of a stochastically accelerated proton.

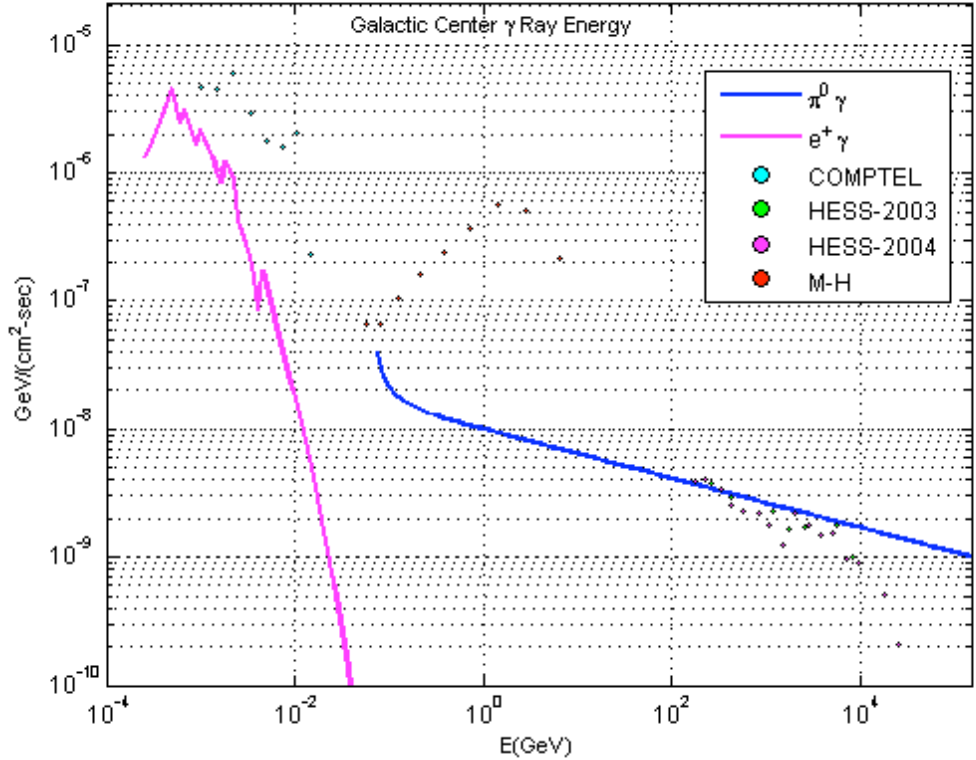

Fig. 1. The magenta curve is our calculated in-flight positron annihilation γ ray spectrum and the blue curve is our calculated $\pi^0$ decay γ ray spectrum. The circles represent data from the Compton satellite [14,16] and HESS array [1,2].

The normalization of the $\pi^0$ decay γ ray spectrum was fixed by assuming that the stochastic acceleration of protons is efficient in the sense that the total flux of stochastically accelerated protons hitting the compact object is comparable to the total flux of protons hitting the object. Since in accordance with what is expected for the processes (2) the protons hitting the surface of the object will produce approximately equal numbers of positrons and $\pi^0$ mesons, we chose to fix the unknown coefficient $\eta F_0$ in eq. (6) by equating the total number of $\pi^0$ decay γ rays to the number of 511 keV positron annihilation γ rays coming from the vicinity of the galatic center. Actually the spectrum (6) diverges near to 70 MeV due to the β factor; in order to calculate the total flux we cutoff the calculation when the $\pi^0$ momentum fell below $Q_0/6$. Obviously our mini-jet model will not be accurate when the momentum of the decay anti-quark is small. Nevertheless the rise in the spectrum as one approaches 70 MeV and sharp cutoff below 70 MeV should be real. Our spectral shape is not consistent with the Egret γ ray data of Mayer-Hasselwander et. al. [14], which shows a peak near to 1 GeV. However, there are hints that the M-H data may be due to a source other than SgA* [15, 16]. At very high energies our mini-jet model should be a good approximation since at high energies the properties of hadronic jets just mimic the properties of the quark or gluon initiating the jet.

Our previous positron spectrum can be used directly to estimate the γ ray production due to in-flight annihilation of positrons with background electrons, since in-flight positron annihilation is mainly important for infalling nucleons with initial energies less than ~ 1 GeV (the cross-section for in-flight annihilation from very energetic positrons is inversely proportional to the energy of the positron). Since the positron production will come mainly from low energy nucleons, we can use our previously calculated Altarelli-Parisi distribution for the momenta of the quarks and gluons inside nucleons at rest to calculate the positron spectrum. The resulting in-flight positron annihilation γ ray spectrum is shown as the magenta curve in Fig 1. The normalization of the positron annihilation curve was chosen to match the 511 keV data, Comparison with EGRET data for the diffuse γ ray background shows that the γ rays resulting from in-flight positron annihilation are probably not observable, in agreement with the result of Beacom and Yuksel [4]. However, it is interesting that overall our theory predicts a sharp dip in the gamma ray spectrum below 70 MeV, which is in accord the EGRET data.

## 3. Conclusion

Our theory provides at least a qualitative explanation for the γ rays which seem to be coming from the galactic center as observed by both the Compton satellite, as the Hess array. Remarkably the total flux of these γ rays seems to be roughly the same as the integrated intensity of positron annihilation radiation coming from the region around the galactic center. Obviously any association of positron annihilation radiation from the galactic center with γ rays coming from the surface of a massive compact object located near Sgr A* would be very exciting, and indeed would provide strong evidence for the idea that compact objects have a physical surface where nucleons decay. Of course, it needs to be kept in mind that the 511 KeV radiation from the galactic center comes from a region very much larger than the size of the compact object near to Sg A*, and it remains to be demonstrated that the observed spatial distribution of this radiation is consistent with positrons with energies of a few MeV diffusing out from a point-like source. Also it may be that the differences between the prediction of our model and the M-H data [14] are due to a failure of our model. Finally, the source of the HESS TeV γ rays may turn out to not be point-like. We eagerly await the results from the GLAST satellite, which may provide a more detailed confirmation of our theory.

**Acknowledgments**

The authors are grateful to Elliott Bloom for suggesting the idea of looking at the effects of cosmic rays hitting dark energy stars and Chris McKee for some helpful comments. This work was partly performed under the auspices of the U.S. Department of Energy by Lawrence Livermore National Laboratory under Contract DE-AC52-07NA27344.

**References**

1. Aharonian, F. A., et. al. 2004, A & A, 425, L13.
2. Aharonian, F. A., et. al. 2006, Phys. Rev. Lett., 97, 221102
3. Barbieri, J. and Chapline, G. 2004, Phys Lett. B, 590, 8.
4. Beacom, J. & Yuksel, Y. 2006, Phys. Rev. Lett., 97, 071102.
5. Chapline, G., Hohlfeld E., Laughlin, R. & Santiago, D. 2001, Phil. Mag. B, 81, 235
6. Chapline, G. 2003, Int J. Mod. Phys. A,, 18, 3587 .
7. T. K. Gaisser, *Cosmic Rays and Particle Physics* (Cambridge University Press 1990).
8. Georgi, H. & Glashow, S. L. 1974, Phys. Rev. Lett., 32, 438 .
9. Hooper, H. and Dingus, B., (2002) astro-ph/0212509.
10. Jean, P., et. al. 2003, Astron. Astrophys. 407, L55.
11. Knodleseder, J., et. al. 2003, Astron. Asrtophys. 411, L457
12. Liu, S. Melia, F., Petrosian, V. & Fatuzzi, M. 2006, ApJ, 647, 1099.
13. Mazur, P. & Mottola, E. 2004, Proc. Nat. Acad. Sci. 111, 9546.
14. Mayer-Hasselwander, H. A. et. al. (1998) , Astron. Asrtophys. 355, 161.
15. Pohl, m. (2004) astro-ph/0412603.
16. Strong, A. W., Moskalenko, I. V. & Reimer, O. 2004, Astron. Astrophys. 444, 495.